\documentclass[runningheads]{llncs}
\usepackage{graphicx}
\usepackage{cite}
\usepackage{booktabs}
\usepackage{algorithm,algorithmic}
\usepackage{hyperref}
\usepackage{subcaption}
\usepackage{amssymb}
\usepackage{pifont}
\usepackage{xcolor}

\definecolor{pink}{RGB}{219, 48, 122}

\begin{document}
\title{Uniformizing Techniques to Process CT scans with 3D CNNs for Tuberculosis Prediction}
\titlerunning{Uniformizing Techniques with 3D CNNs for Tuberculosis Prediction}

\author{Hasib Zunair\inst{1} \and
Aimon Rahman\inst{2} \and
Nabeel Mohammed\inst{3} \and \\
Joseph Paul Cohen\inst{4}\\}
%
%
\institute{Concordia University\inst{1}, North South University\inst{2,3} and Mila, University of Montreal\inst{4}\\
\email{hasibzunair@gmail.com}\inst{1},
\email{irsnigdha@gmail.com}\inst{2}, \email{nabeel.mohammed@northsouth.edu}\inst{3}
\email{joseph@josephpcohen.com}\inst{4}}
\maketitle              
\begin{abstract}

A common approach to medical image analysis on volumetric data uses deep 2D convolutional neural networks (CNNs). This is largely attributed to the challenges imposed by the nature of the 3D data: variable volume size, GPU exhaustion during optimization. However, dealing with the individual slices independently in 2D CNNs deliberately discards the depth information which results in poor performance for the intended task. Therefore, it is important to develop methods that not only overcome the heavy memory and computation requirements but also leverage the 3D information. To this end, we evaluate a set of volume uniformizing methods to address the aforementioned issues. The first method involves sampling information evenly from a subset of the volume. Another method exploits the full geometry of the 3D volume by interpolating over the z-axis. We demonstrate performance improvements using controlled ablation studies as well as put this approach to the test on the ImageCLEF Tuberculosis Severity Assessment 2019 benchmark. We report 73\% area under curve (AUC) and binary classification accuracy (ACC) of 67.5\% on the test set beating all methods which leveraged only image information (without using clinical meta-data) achieving 5-th position overall. All codes and models are made available at \href{https://github.com/hasibzunair/uniformizing-3D}{\texttt{\textcolor{pink}{https://github.com/hasibzunair/uniformizing-3D}}}.


\end{abstract}
\section{Introduction}
To learn the geometric properties of volumetric data, there are challenges imposed by the data itself \cite{ahmed2018survey}. One major challenge is fitting the data in GPU memory during optimization. Furthermore, complicacy also arises when dealing with the variable depth size in the data. Hence, the data preparation scheme plays a vital role to build robust systems comprising volumetric image data. In the context of medical imaging, deep learning \cite{ref_14} has been widely used in a variety of tasks and domains \cite{ref_cad, ref_3, ref_23, ref_6, ref_83}. While many of these medical image modalities are two dimensional (2D), computed tomography (CT) volumes are three dimensional (3D) and require more computational expense which can be an insurmountable obstacle in the case of limited GPU memory.

This necessitates the use of 2D CNN architectures \cite{ref_7,ref_8,UCSD,21-ref,24,26,2d-bad} where the 3D data is treated as a set of independent \textit{slices}. However, there is evidence that better results are achievable when using the full volumetric data \cite{bad-1, bad-2, bad-3, ref_8,3d-good}. An observation is that 2D approaches discard information along the depth dimension/z-axis and prevents to preserve 3D context~\cite{bad-4} leading to non-optimal performance. On the other hand, memory challenges are also experienced due to the nature of 3D data. It is also noteworthy to mention that 3D datasets exist which are annotated at the slice level, where it is justified to use 2D CNN approaches~\cite{ref_7} but it is not the case when the data is annotated at the volume level \cite{ref_15, ref_16}.

In this work, we evaluate a set of volume uniformizing methods in the 3D image domain. First, we explore sampling a subset of the slices using a spacing factor to evenly sample from the sequence of slices to construct the desired volumetric output. However, deliberately losing information likely prevents learning robust representations from the data with the risk of adding artifacts. We explore interpolating over the z-axis to capture information from multiple slices which turns out to be a very reasonable solution and provides good performance and at the same time satisfy GPU memory requirements. We put our technique to the test using 3D medical images originating from the Computed Tomography (CT) domain with annotations at the CT/volume level. This is evaluated on the ImageCLEF Tuberculosis Severity Assessment 2019 test set which outperforms all methods leveraging only image information and achieves 5-th position overall. We summarize our contributions as follows:

\begin{enumerate}
  \item We evaluate Even Slice Selection (ESS) and Spline Interpolated Zoom (SIZ) which exploit the full geometry of the 3D CT data based on improvements upon SSS \cite{ref_imageclef_2019}.
  \item We develop a $17$-layer 3D convolutional neural network inspired by \cite{maturana2015voxnet} with major modifications.
  \item We perform controlled ablation studies and show superior performance and reliability for SIZ, both qualitatively and quantitatively.
  \item We evaluate our best approach on the ImageCLEF Tuberculosis Severity Assessment 2019 benchmark which outperforms all methods leveraging only image information achieving 5-th position overall.
\end{enumerate}

\section{Related Work}

\noindent{\textbf{2D Approaches.}}\quad To mimic the 3-channel image representation (i.e.,  RGB), prior works follow multi-slice representation of 3D images as 2D inputs \cite{gerard2020multi}. UUIP\_BioMed \cite{biomed} proposes a CNN using 2D projections of 3D CT scans which provide a probability score. HHU \cite{18-ref} demonstrates a multi-stage approach where they first assess the CT-findings for another task and then apply linear regression to obtain the final predictions. A hybrid 2D CNN was trained by creating 2D derived images by concatenating sagittal and coronal CT slices by CompElecEngCU \cite{22}. Ensembling of 2D CNNs was also demonstrated by SD VA HCS/UCSD \cite{UCSD} for tuberculosis (TB) prediction. MedGIFT \cite{20-ref} used a graph-based approach by dividing the lung fields into several subregions (different for each patient) and considered these subregions as nodes of a graph. These graphs were transformed into lung descriptor vectors and then classified. UniversityAlicante \cite{23} proposed to use each CT volume as a time series and used optical flow on the 3 directions. MostaganemFSEI \cite{21-ref} first selected relevant axial CT slices
manually. Features are extracted using a 2D CNN which is followed by a long short term memory (LSTM) to obtain the final predictions. SSN CoE \cite{24} manually selected a subset of slices for each patient and then used a 2D CNN for classification. FIIAugt \cite{26} performed a random sampling of pixels of the CT volumes and used a combination of decision trees and weak classifiers.

\medskip
\noindent{\textbf{3D Approaches.}}\quad Instead of regarding the 3D spatial information as the input channel in 2D based methods, studies based on 3D convolutions for 3D medical image analysis have been demonstrated \cite{yang2019reinventing, gordaliza2019multi,wu2019deep}.
These methods are capable of capturing the 3D context in any axis and mitigates the limited 3D context along a certain axis (depth/z-axis) in 2D approaches. Hence, the 3D methods are generally better when the 3D context is required (e.g, locating nodules). A related study is UUIP \cite{UUIP}, where they use 3D CNN as an autoencoder followed by a random forest classifier. Before training the autoencoder, the downsampling was performed at the volume level to preserve the 3D context. UoAP \cite{ref_imageclef_2019} used a 3D CNN (VoxNet) with either 16 or 32 slices that were selected from the top, middle, and bottom of the volume.

\section{Methods}
In this section, we describe the main components and the algorithmic steps of the methods employed for the task of TB prediction. Our goal is to learn a discriminative function $f(\textbf{X}) \in  \{\textbf{0},\textbf{1}\}$, where 1 indicates high TB severity and 0 otherwise. $\textbf{X}$ represents a CT  scan volume of size $W\times H\times D$, where W, H, and D represent the width, height, and depth of the volume respectively.
\subsection{Uniformizing Techniques} \label{techniques}
We discuss the techniques in detail which we use to prepare the data before learning $f(\textbf{.})$, the discriminative function. We talk about the algorithmic step of each technique and show qualitative results. It is important to mention that readers should not be confused when we refer to the term \textit{slices}, it means that the sampling is done at the slice level to acquire the desired output \textit{volume}. 

\begin{figure}
\centering
\includegraphics[scale=.4]{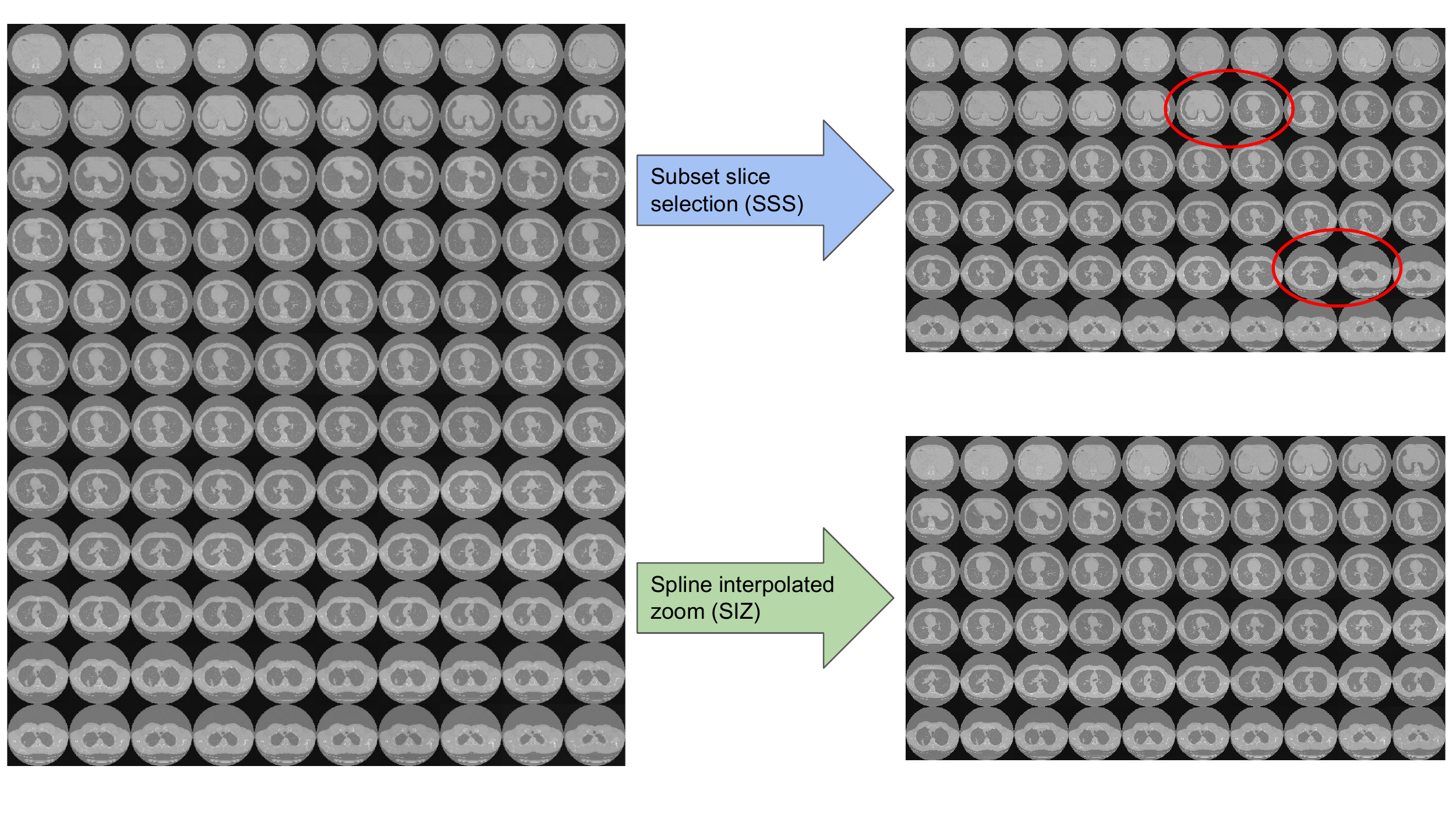}
\caption{\small Visualization of raw CT slices from an arbitrary 3D CT scan from ImageCLEF Tuberculosis Severity Assessment 2019 (left), qualitative comparison of SSS and SIZ (right) which shows SSS changes the semantic meaning of the volumetric data as a subset of the \textit{slices} are being discarded, variations marked in red circles. This results in information loss where SIZ on the other hand maintains.}
\label{vis}
\end{figure}

\subsubsection{Slice Selection} 
\label{ss}
Depth variability of the 3D CT scans motivate the concept of sampling from \textit{slice} level to construct the desired \textit{volume} and balance between model performance and GPU memory constraints~\cite{ref_7, ref_8, ref_imageclef_2019, bad-1, bad-2, bad-3}.

\vspace{-1cm}
\textit{Subset Slice Selection} (SSS): In this technique originally proposed in \cite{ref_imageclef_2019}, \textit{slices} are sampled from the first, middle and last position of the entire volume. The middle slices are sampled by indexing from half of the input volume depth to ensure consistency due to the depth variability. A depthwise stack is then performed over the subsets to attain the desired input volume. An illustration is shown in Fig.\ref{vis}.  

\textit{Even Slice Selection} (ESS): A major drawback of SSS is that it prevents us from using the remaining subset of the data which causes the processed volume not to be representative of the original volume. We show qualitative evidence in Fig.~\ref{vis} which is indicated by red circles. ESS can be considered as an improved version of SSS which provides good performance compared to SSS. In ESS, a target depth $N$ and a scan depth of size $D$ is computed. A spacing factor is then determined by the equation $F = \frac{D}{N}$. Sampling is done at the slice level by maintaining the spacing factor $F$ between the sequence of slices in the volumetric data. This gives a better representation compared to the SSS technique as we show experimentally in later sections. The algorithmic steps are shown in Algorithm~\ref{algo:algoooo1}.
\vspace{-0.5cm}
\subsubsection{Spline Interpolated Zoom (SIZ)}
Even though ESS preserves the representation to an extent, the desired volume is still acquired from a subset of the data. Therefore, to discard the concept of sampling from independent \textit{slices}, an alternative solution is Spline Interpolated Zoom (SIZ) which enables even better representation of the volumetric data. Similar techniques have been used in the other studies \cite{UUIP, yang2019reinventing, gordaliza2019multi,wu2019deep}. In this technique, instead of manually selecting a subset of slices, a constant target depth size of $N$ is pre-determined. We then take each volume, calculate its depth $D$, and zoom it along the z-axis by a factor of $\frac{1}{D/N}$ using spline interpolation \cite{spline}, where the interpolant is an order of three.  Here, the input volume is zoomed or squeezed by replicating the nearest pixel along the depth/z-axis. A visual representation of this can be seen in Fig.\ref{vis} along with the algorithm summarized in Algorithm~\ref{algo:algoooo2}. As it uses spline interpolation to squeeze or expand the z-axis to the desired depth, it retains a substantial level of information from original 3D volume as opposed to the aforementioned techniques, SSS and ESS, which discarded a part of the volumetric data.

\noindent \begin{minipage}{0.44\textwidth}
\begin{algorithm}[H]
    \centering
    \caption{Even Slice Selection (ESS) data processing method}
  \label{algo:algoooo1}
  \begin{algorithmic}[1]
  \small
    \REQUIRE A 3D volumetric image $I$ of size $W\times H\times depth$.
    \ENSURE $I$ is a rank 3 tensor.
    \STATE Set constant target depth of size $N$ 
    \STATE Compute $depth$ denoted as $D$
    \STATE Compute depth factor by $F = \frac{D}{N}$
    \STATE Sample slices from the volume by maintaining the depth factor $F$
    \STATE Output processed volume $I'$ of dimension $W\times H\times N$
  \end{algorithmic}
\end{algorithm}
\end{minipage}
\hfill
\begin{minipage}{0.52\textwidth}
\begin{algorithm}[H]
    \centering
    \caption{Spline Interpolated Zoom (SIZ) data processing method}
  \label{algo:algoooo2}
  \begin{algorithmic}[1]
  \small
    \REQUIRE A 3D volumetric image $I$ of size $W\times H\times depth$.
    \ENSURE $I$ is a rank 3 tensor.
    \STATE Set constant target depth of size $N$ 
    \STATE Calculate the $depth$ denoted as $D$
    \STATE Compute depth factor by $\frac{1}{D/N}$ denoted as $DF$ 
    \STATE Zoom $I$ using spline interpolation \cite{spline} by the factor $DF$ 
    \STATE Output processed volume $I'$ of dimension $W\times H\times N$
  \end{algorithmic}
\end{algorithm}
\end{minipage}

\subsection{Three-dimensional (3D) CNN architecture} \label{B}
\begin{figure}
\centering
\includegraphics[width=0.95\textwidth]{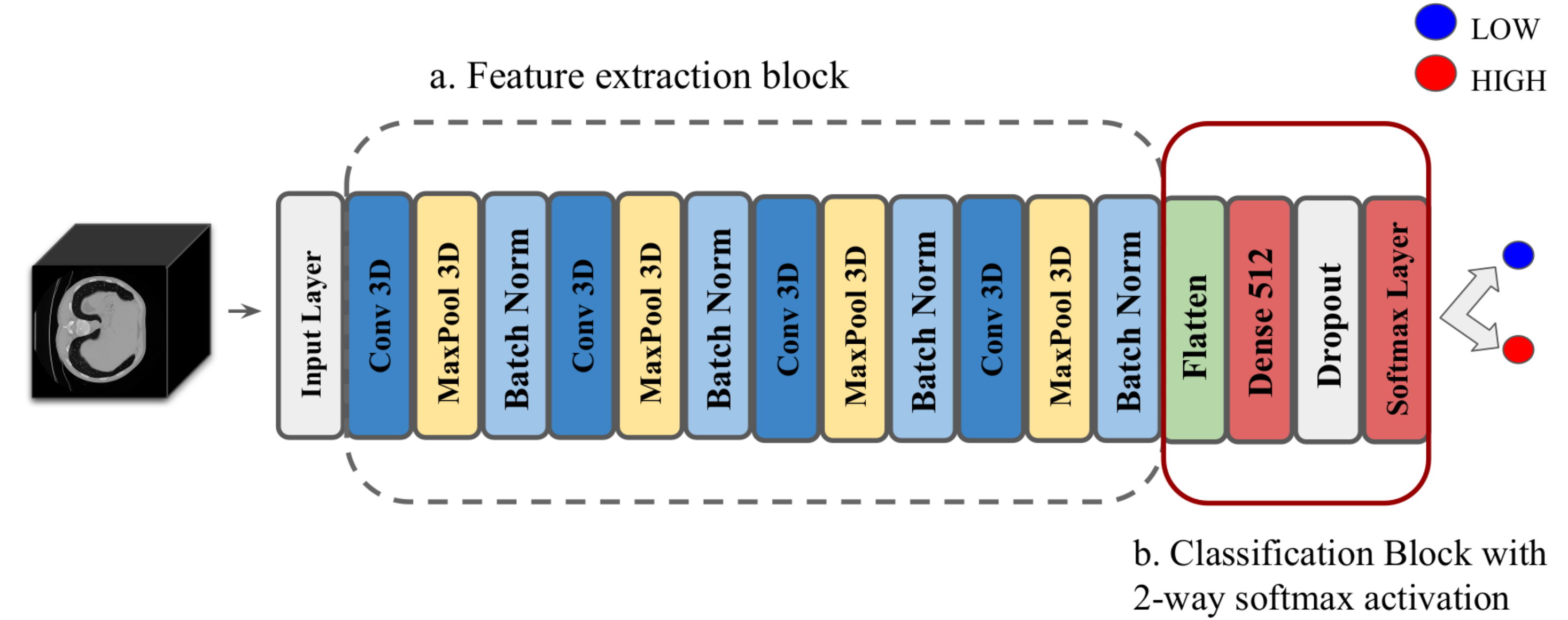}
\vspace{1pt}
\caption{\small Our proposed $17$-layer 3D convolutional neural network architecture which consist of several modules of 3D conv, maxpool and batch normalization layers.} \label{archi}
\end{figure}

Inspired from \cite{ref_voxnet}, we design a $17$ layer 3D CNN which comprises four 3D convolutional (CONV) layers with two layers consisting of $64$ filters followed by $128$ and $256$ filters all with a kernel size of $3\times 3 \times 3$. Each CONV layer is followed by a max-pooling (MAXPOOL) layer with a stride of $2$ and ReLU activation which ends with batch normalization (BN) layer~\cite{batch_norm}. Essentially, our feature extraction block consists of four CONV-MAXPOOL-BN modules. The final output from the feature extraction block is flattened and passed to a fully connected layer with 512 neurons. We use an effective dropout rate of 60\% similar to~\cite{pattnaik2019predicting}. Due to a coding error, we implement this using two dropout layers~\cite{drop}. The output is then carried to a dense layer of 2 neurons with softmax activation for the binary classification problem. The network architecture is shown in Figure \ref{archi}. 

We consider keeping the network relatively simple to avoid overparameterization~\cite{raghu2019transfusion} problems with only 10,658,498 learnable parameters. This is also motivated by the fewer number of training samples and the memory challenges associated with it.

\section{Experiments}

\subsection{Dataset}
The dataset is provided by ImageCLEF Tuberculosis 2019 \cite{ref_15}\cite{ref_16}, intended for the task of severity scoring (SVR). It consists of a total 335 chest 3D CT scans with annotation of high and low provided by a medical doctor and also lung segmentation masks, in addition to clinically relevant metadata was also available which includes the following binary measures: disability, relapse, symptoms of TB, comorbidity, bacillary, drug resistance, higher education, ex-prisoner, alcoholic, smoking. From the dataset, 218 individual chest CT scans are provided for training, and the remaining 117 were held out for the final evaluation in the ImageCLEF evaluation platform. The CT images each have a dimension of $512\times $512 pixels and the depth size varying from about 50 to 400 which store raw voxel intensity in Hounsfield units (HU). Figure~\ref{vis} shows an instance of this.

\vspace{-0.4cm}
\subsection{Baseline and Implementation details}
We consider SSS~\cite{ref_imageclef_2019} as the baseline. Each configuration embodies a different type of processing discussed in Section~\ref{techniques}. All configurations are based on the network described in Section~\ref{B} and trained on a machine with NVIDIA 1050Ti with 4GB memory. We used Stochastic Gradient Descent(SGD) optimizer with a learning rate of $10^{-6}$ and a momentum of $0.99$, parameters found from multiple trials. Weight is initialized using the Glorot initialization method \cite{glorot2010understanding} and minimize the Mean average error \cite{mae} during training. During training, the network accepts input of size $128\times 128 \times 64$ with a batch size of 2. We tried increasing the depth size to more than 64 but resulted in GPU memory error. For our experiments with ESS, we found four CT scans which had a depth of less than 64 with a minimum being $47$. In these cases, we first apply ESS and then calculate the difference with the target depth, $64$, and repeatedly add the last slice until the target depth is reached. We resize to $128\times 128$ on the slice level and then use techniques discussed in Section \ref{ss} to get the desired volume. To ensure a fair comparison between the uniformizing methods, we keep the desired input size of $128\times128\times 64$ for all our experiments. We provide code and model to reproduce our experiments at \href{https://github.com/hasibzunair/uniformizing-3D}{\texttt{\textcolor{pink}{https://github.com/hasibzunair/uniformizing-3D}}}.

\subsection{Metrics} \label{D}
As per challenge rules, the task is evaluated as a binary classification problem. The evaluation metrics are Area Under the ROC Curve (AUC) and accuracy (ACC), prioritizing the former by the challenge organizers. We refrain from using other evaluation metrics since it would limit our comparison with the approaches proposed in the challenge.

\subsection{Results}
In this section, extensive experiments are conducted to evaluate the performance of the uniformizing methods. First, we compare our methods with the baseline on the ImageCLEF Tuberculosis Severity Assessment 2019 benchmark. Since the dataset is small, we also perform cross-validation tests to estimate the general effectiveness and reliability of the methods, ensuring a balance between bias and variance. Finally, we show ablations of orthogonal preprocessing and how our method performs related to other methods on the ImageCLEF Tuberculosis Severity Assessment 2019 benchmark.

\begin{figure}
\includegraphics[scale=.6]{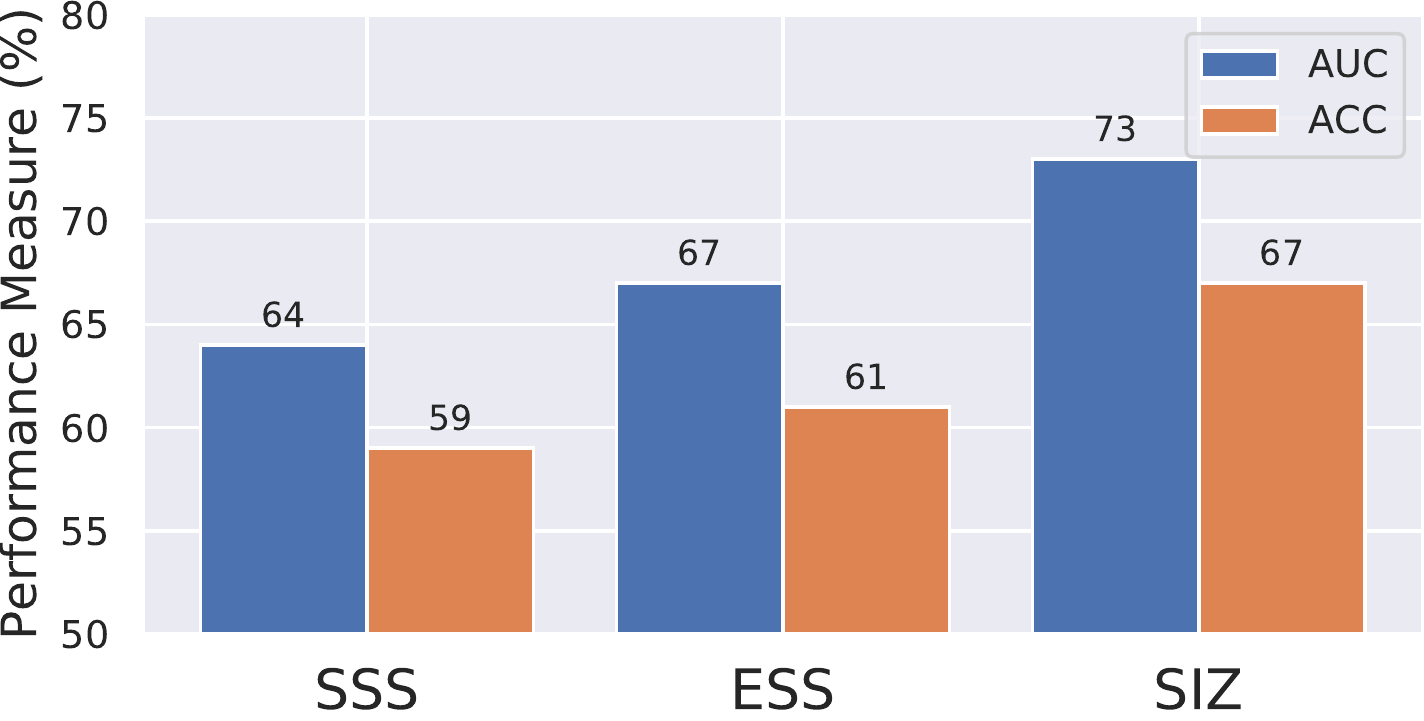}
\centering
\caption{\small Performance measures reported are evaluated on the test set provided by ImageCLEF Tuberculosis Severity Assessment 2019 consisting of 117 chest 3D CT scans.}
\label{res_fig}
\end{figure}

\medskip
\noindent{\textbf{Comparison with baseline.}}\quad We believe SIZ better represents the 3D CT when downsampled compared to SSS and ESS. This is depicted in Figure~\ref{res_fig}, which shows that SIZ yields better performance in both metrics by a margin of $9\%$ and $8\%$ compared to SSS and is of significance. We further validate this by showing the qualitative comparison in Figure~\ref{vis} in which we show visual evidence that slice selection methods do not leverage information from full 3D CT scans which SIZ does. It is also observed that ESS yields slightly better results than SSS. This is because even though ESS samples from half of the volume, the sampling is carried out in a sequential process. This approach results in a better representation of the 3D CT scan compared to SSS where a subset of slices is sampled from predefined points. Thus, selecting specific \textit{slices} does not preserve the semantic meaning of volumetric data as it is not the proper representation of the 3D CT scan which is also intuitive. Even though ESS is downsampling the volume from a subset, this still results in better performance as the sampling is done throughout the entire volume. In particular, ESS increases the probability of sampling the TB affected slices compared to SSS. Since TB infection can affect any part of the lung, it is also not possible to determine which \textit{slices} are to be discarded without looking at the scans individually. As the annotations are provided at the volume level and not at the slice level, it is crucial to retrieve information from the entire volume. 

\begin{figure}
\begin{subfigure}{.48\textwidth}
  \centering
  \includegraphics[width=1.0\linewidth]{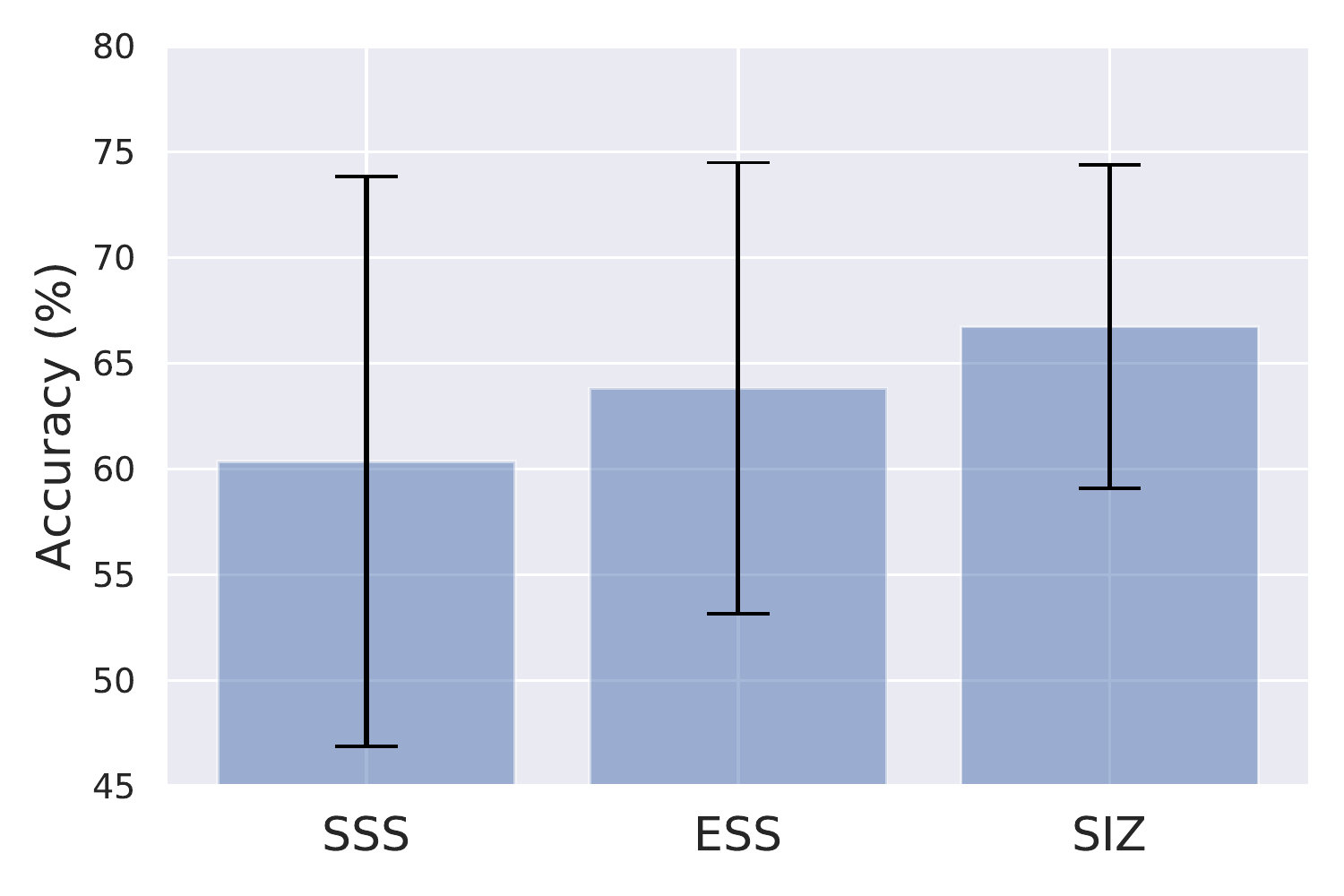}
  \caption{Mean accuracy and standard deviation of cross validation results over 10 trials with randomly separated 80\% training and 20\% test images.}
  \label{fig:1a}
\end{subfigure}%
\hspace{0.2cm}
\begin{subfigure}{.45\textwidth}
  \centering
  \includegraphics[width=1.0\linewidth]{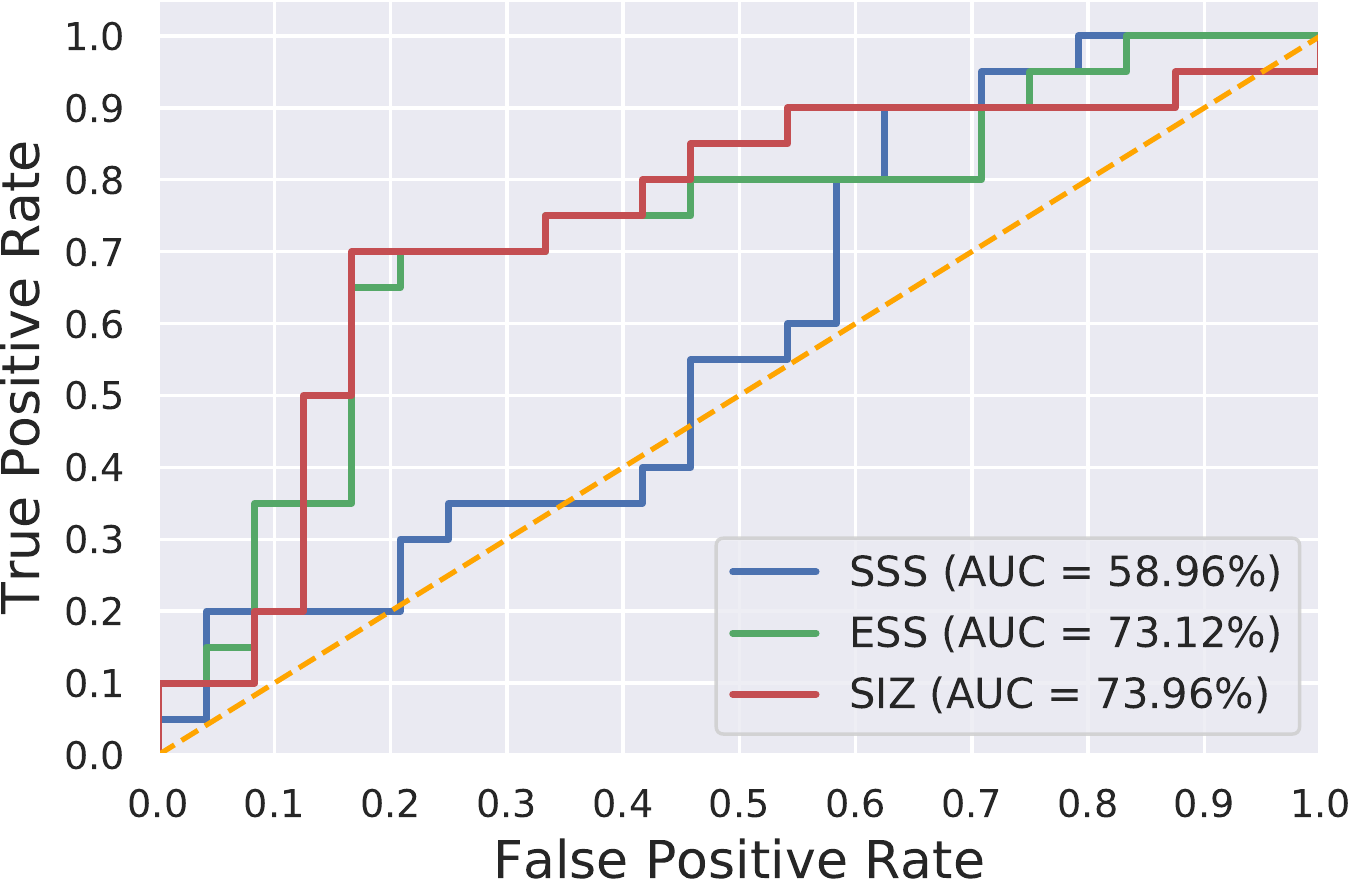}
  \caption{ROC curves for SSS, ESS and SIZ, along with the corresponding AUC values from best weights for each configuration.}
  \label{fig:1b}
\end{subfigure}
\caption{Performance measures reported from 10 runs of the cross-validation.}
\label{fig:fig}
\end{figure}

\medskip
\noindent{\textbf{Cross validation.}}\quad We also report the cross-validation results as shown in Figure~\ref{fig:1a}. It can be seen that SIZ not only has a higher mean accuracy than the baselines but also has a lower standard deviation owing to more reliability. Figure~\ref{fig:1b} displays the ROC curve, which shows better performance of the method SIZ compared to the baselines. It is to be noted that the ROC curves reported are from the best performing results on the validation set. Each point on ROC represents a different trade-off between false positives and false negatives. A ROC curve that is closer to the upper right indicates better performance (TPR is higher than FPR). Even though during the early and last stages, the ROC curve of SIZ seems to highly fluctuate at certain points, the overall performance is much higher than the baselines, as indicated by the AUC value. This better performance demonstrates that 3D context plays a crucial role and enables the model to learn effective representations.

\noindent{\textbf{Ablation study.}}\quad Table~\ref{table:orthogonal} illustrates the ablations on orthogonal preprocessing. For all configurations SSS, ESS, and SIZ we observe performance improvements on both AUC and ACC after pixel normalization and zero-centering. Since the 3D CT scans have raw voxel intensities in Hounsfield units (HU), we normalize the values between [0,1]. We then perform zero-centering by subtracting the total mean value from each pixel, making the mean of the total dataset zero. For SIZ, the increase in performance compared to baseline is the larget with an increase of $11\%$ and $14\%$ margin in AUC and ACC respectively.

\begin{table}[t]
\setlength{\tabcolsep}{4pt}
\centering
\caption{Ablations of orthogonal preprocessing evaluated on the final test set provided by ImageCLEF Tuberculosis Severity Assessment 2019 consisting of 117 chest CT scans.}
\begin{tabular}{l c c c c}
\toprule
\bfseries Method & \bfseries Normalize & \bfseries Zero Center & \bfseries AUC & \bfseries ACC\\
\midrule
SSS   & No & No &   0.626 & 0.538\\
SSS   & Yes & No &   0.635 & 0.573\\
SSS   & Yes & Yes &   0.640 & 0.598\\
ESS   &   No & No   &  0.639 & 0.607\\
ESS   &   Yes  & No   & 0.667 & 0.598\\
ESS   &   Yes  & Yes   & 0.670 & 0.611\\
SIZ   & No & No &  0.648 & 0.581\\
SIZ   & Yes & No &  0.652 & 0.607\\
SIZ   & Yes & Yes &  \textbf{0.730} & \textbf{0.675}\\
\bottomrule
\end{tabular}
\label{table:orthogonal}
\end{table}

\medskip
\noindent{\textbf{ImageCLEF Tuberculosis Severity Assessment 2019 Benchmark.}}\quad We summarize the results in Table \ref{tab:comp} which report the performance evaluation on the final test set. It is observed that our best method, SIZ, achieves comparable performance with the top-ranking methods. It is noteworthy to mention that UUIP\_Biomed \cite{biomed}, UUIP\cite{UUIP}, HHU \cite{18-ref} and CompElecEngCU \cite{22} leverage the clinically relevant metadata in order to significantly improve performance and also develop multi-stage approaches which adds complexity~\cite{ref_15}. We increased the input volume to the $128\times128\times128$, the same as UUIP\cite{UUIP} which results in a model almost three times larger than ours with 29,532,866 learnable parameters and led to a memory error. Even with using only image information, our method performs better than SD VA HCS/UCSD \cite{UCSD} where they used an ensemble of 2D CNNs and the relevant meta-data. It also performed better than the 3D CNN method by UoAP \cite{ref_imageclef_2019}. Our best method also outperforms several 2D approaches such as MedGIFT \cite{21-ref}, SSN CoE \cite{24} and FIIAugt \cite{26}.

From Table \ref{tab:comp} it is also seen that among the top-ranking results which only use image information (no meta-data), our method achieves the best results. Even though MedGIFT \cite{20-ref} did not use any meta-data, they were the only team that used the lung segmentation masks.

\begin{table}[t]
\centering
\setlength{\tabcolsep}{1pt}
\caption{\small Performance metric results compared with previous top ranking approaches on ImageCLEF Benchmark. The results reported on each of the metrics are on the ImageCLEF test set which consists of 117 3D CT scans. Boldface indicates our best method}
\begin{tabular}{l c c c c c}
\toprule
\bfseries Group Name & \bfseries Method Type & \bfseries Input Volume & \bfseries AUC & \bfseries ACC & \bfseries Meta-data\\
\midrule
UIIP\_BioMed \cite{biomed} & 2D & None & 0.7877 & 0.7179  & Yes \\
UIIP \cite{UUIP} & 3D - None & $(128\times128\times128)$ & 0.7754 & 0.7179 & Yes \\
HHU \cite{18-ref} & 2D & None &  0.7695 & 0.6923 & Yes \\
CompElecEngCU \cite{22} & 2D & None & 0.7629 & 0.6581 & Yes \\
\textbf{Ours} & 3D - SIZ & $(128\times128\times64)$ & \textbf{0.7300} & \textbf{0.6750} & No \\
SD VA HCS/UCSD \cite{UCSD} & 2D & None & 0.7214 & 0.6838 & Yes \\
MedGIFT \cite{20-ref} & 2D & None & 0.7196 & 0.6410 & No \\
UniversityAlicante \cite{23} & 2D & None & 0.7013 & 0.7009 & No \\
MostaganemFSEI \cite{21-ref} & 2D & None & 0.6510 & 0.6154 & No \\
SSN CoE \cite{24} & 2D & None & 0.6264 & 0.6068 & No \\
UoAP \cite{ref_imageclef_2019} & 3D - SSS & $(128\times128\times32)$ & 0.6111 & 0.6154 & No \\
FIIAugt \cite{26} & 2D & None & 0.5692 & 0.5556 & No \\
\bottomrule
\end{tabular}
\label{tab:comp}
\end{table}

\section{Discussion and Conclusion}

We address the problem of variable volume size and heavy computation requirements during optimization when dealing with 3D image data. In particular, we evaluate a set of volume uniformizing methods applied to 3D medical images in the CT domain for the task of TB prediction. We hypothesize that analyzing 3D images in a per slice (2D) basis is a sub-optimal approach that can be improved by 3D context if computational challenges can be overcomed. We systematically evaluate different ways of uniformizing CT volumes so that they fit into memory and determine interpolating over the z-axis to be the best.
We further validate this approach on the ImageCLEF benchmark obtaining 5th place and beat all methods which operate on the CT image alone without patient metadata.

%
%
%
\bibliographystyle{splncs04}
 \bibliography{1_reference}

\end{document}